\documentstyle[preprint,aps]{revtex} % For drafts and producing CLNS reports

\begin{document}

\draft % makes pacs numbers print
% Begin deletion for submission to PRL or PRD
\tighten % To save trees, please use this for the copy to be distributed!
\preprint{\vbox{\hbox{CLNS 94/1286  \hfill}
                \hbox{CLEO 94--16 \hfill}
                \hbox{\today       \hfill}}}
% End   deletion for submission to PRL or PRD

\title{
Semileptonic Branching Fractions of Charged and Neutral $B$ Mesons
}

%Author list here
\author{
M.~Athanas,$^{1}$ W.~Brower,$^{1}$ G.~Masek,$^{1}$ H.P.~Paar,$^{1}$
J.~Gronberg,$^{2}$ R.~Kutschke,$^{2}$ S.~Menary,$^{2}$
R.J.~Morrison,$^{2}$ S.~Nakanishi,$^{2}$ H.N.~Nelson,$^{2}$
% ...
}

\address{
\bigskip % Delete for submission to PRL or PRD
{\rm (CLEO Collaboration)}\\  % DO NOT Delete!
\newpage % Delete for submission to PRL or PRD
$^{1}${University of California, San Diego, La Jolla, California 92093}\\
$^{2}${University of California, Santa Barbara, California 93106}\\
% ...
\bigskip % Delete for submission to PRL or PRD
}        % DO NOT delete!
%\date{\today}
\maketitle

\begin{abstract}
An examination of leptons in ${\Upsilon (4S)}$ events tagged by reconstructed
$B$ decays yields semileptonic branching fractions of
$b_-=(10.1 \pm 1.8\pm 1.4)\%$
for charged and
$b_0=(10.9 \pm 0.7\pm 1.1)\%$
for neutral $B$ mesons.
This is the first measurement for charged $B$.
Assuming equality of the charged and neutral semileptonic widths,
the ratio $b_-/b_0=0.93 \pm 0.18 \pm 0.12$ is equivalent to the ratio
of lifetimes.
\end{abstract}
\pacs{14.40.Nd, 13.20.Jf}

\narrowtext % This is for the narrow columns in galley - not active in preprint

%
%============================INTRODUCTION==============================
%
\vskip 15pt
Semileptonic $B$ decay proceeds through a coupling of the $b$-quark
to a $c$- or $u$-quark and a $l^-{\bar{\nu}}$ pair (via a virtual $W^-$).
This is known as a {\it spectator} process because the accompanying quark
plays no direct role.
In the na\"{\i}ve spectator model for hadronic decays, the relationship
between semileptonic and hadronic widths, and thus the semileptonic
branching fraction, is readily predicted\cite{bigi}.
Unfortunately, this simple picture does not seem to hold;
for the past decade, the measured average branching fraction,
$\bar b$, of $B$'s in ${\Upsilon (4S)}$ events \cite{bargus,bcleo,rwang}
has been $\sim$15\% lower than predictions\cite{bigi}.

The hadronic width may be altered by contributions from nonspectator
diagrams or by interference among final state quarks in spectator
processes.
Both mechanisms carry intrinsic dependences on the flavor of the spectator
and do not apply to semileptonic decays.
They may therefore result in unequal charged and neutral semileptonic
branching fractions, $b_-$ and $b_0$.
The asymmetry is predicted to be less than 10\%\cite{bigi}.

Previous measurements in $\Upsilon(4S)$ events carried
several uncertainties.
If the $\Upsilon$(4S) decays to final states other than ${B\bar{B}}$,
there will be an apparent shift in $\bar b$\cite{theo}.
Rates of exclusive decay determine
$(b_-f_{+-})/(b_0f_{00})$, where $f_{+-}$ and $f_{00}$ are the
production fractions of neutral and charged ${B\bar{B}}$
events\cite{RMEASURED}.
This gives $b_-/b_0$ if ${f_{+-}/{f_{00}}}=1$, but the uncertainty on
this assumption is a major source of systematic error.
These uncertainties are minimized in {\it tagged} measurements
where the number of $B$ mesons in the event sample is counted directly
rather than being inferred from the number of ${\Upsilon (4S)}$ events.
We report here the measurements via tagging of $b_-$, $b_0$ and their ratio.
This is a first measurement of $b_-$.
If there is no asymmetry in semileptonic widths, then
the ratio $b_-/b_0$ is equal to the ratio of lifetimes, which is
measured at higher energies \cite{LEP}.

%
%=============================Data Sample==============================
%

The data were collected with the CLEO
II detector\cite{detector} at the Cornell Electron Storage Ring (CESR) and
consist of integrated luminosities of 1.35~fb$^{-1}$ on the ${\Upsilon (4S)}$
resonance and 0.64~fb$^{-1}$ taken at a CM energy which is
lower by 60~MeV (continuum).
All events considered in this report are required to pass our standard
hadronic criteria, which require at least 3 well-fitted charged tracks, a
measured energy at least 0.15 times the CM energy and
an event vertex consistent with the known interaction point.

$B$ decays are reconstructed using three methods,
(A) full reconstruction of all decay products, (B) partial reconstruction
of semileptonic decay, and (C) partial reconstruction of a
hadronic decay.
Tag (A) gives the only measurement of $b_-$.
All three yield $b_0$, where method (B) dominates statistically.
Each analysis is performed both with and without a requirement that events
contain a hard lepton consistent with being a primary decay product
of the other $B$.
For ${B^-}$, we need only consider leptons with charge corresponding
to ${B^+}$ \cite{cconj}.
For $\bar{B}^0$, since $B^0$ mixes with $\bar{B}^0$, we accept leptons of
either sign.

%=============================Tagging methods==========================
%
%
%===============================B Reconstr.============================
%
In tag (A), we reconstruct hadronic $B$ decays \cite{CABS}
in eight modes: $D {\pi^{-}}$, ${D^{\ast}}{\pi^{-}}$, $D {\rho^{-}}$,
${D^{\ast}}{\rho^{-}}$, $D {a_1^-}$, ${D^{\ast}}{a_1^-}$, $\psi K$ and
$\psi {K^{\ast}}$, with charm mesons in the channels:
\begin{center}
\begin{tabular}{l l l}
${D^{\ast+}}$ & ${\rightarrow}$ & ${D^{0}}{\pi^{+}}$, ${D^{+}}{\pi^{0}}$     \\
${D^{\ast0}}$ & ${\rightarrow}$ & ${D^{0}}{\pi^{0}}$  \\
${D^{0}}$  & ${\rightarrow}$ & ${K^{-}}{\pi^{+}}$, ${K^{-}}{\pi^{+}}{\pi^{0}}$,
${K^{-}}{\pi^{+}} {\pi^{+}}{\pi^{-}}$,
${K^0_{\rm S}} {\pi^{0}}$, ${K^0_{\rm S}}{\pi^{+}}{\pi^{-}}$ \\
${D^{+}}$  & ${\rightarrow}$ & ${K^{-}}{\pi^{+}}{\pi^{+}}$, ${K^0_{\rm
S}}{\pi^{+}}$ \\
$J/\psi$& ${\rightarrow}$ & ${e^{+}}{e^{-}}$, ${\mu^{+}}{\mu^{-}}$    \\
\end{tabular}
\end{center}
and light mesons reconstructed in the channels
${\pi^{0}} {\rightarrow} {\gamma}{\gamma}$, ${K^0_{\rm S}}{\rightarrow}
{\pi^{+}}{\pi^{-}}$,
${K^{\ast +}} {\rightarrow} K^+ \pi^0$,
${K^{\ast +}} {\rightarrow} K^0_{\rm S} \pi^+$,
${K^{\ast 0}} {\rightarrow} K^0_{\rm S} \pi^0$,
${K^{\ast 0}} {\rightarrow} K^+ \pi^-$,
${\rho^{+}}{\rightarrow} {\pi^{0}}{\pi^{+}}$, ${\rho^{0}}{\rightarrow}
{\pi^{+}}{\pi^{-}}$ and
$a_1^+ {\rightarrow} {\rho^{0}}{\pi^{+}}$.

All events must contain at least four well-fitted
charged tracks and have a ratio $R_2$ of second and zeroth Fox-Wolfram
moments\cite{R2} less than $0.45$.
Leptons are required to have momentum $1.4$--2.4~GeV,
to be within the barrel region of the detector,
and to be consistent with originating at the interaction point.
Muons must penetrate at least five
nuclear absorption lengths in the muon detector.
Electron identification utilizes primarily
the ratio of calorimetric energy to momentum and
specific ionization in the drift chamber.
Except in the case of direct slow pions from ${D^{\ast}}$,
both the time-of-flight and $dE/dx$ systems are used
for hadron identification.
Photons must be detected in regions of good calorimeter resolution and
exceed a minimum energy, equal to 30~MeV for most of the detector
acceptance.
Candidates for ${\pi^{0}}$, ${K^0_{\rm S}}$, $D$, ${D^{\ast}}$,
and $J/\psi$ are required to have an
invariant mass (for ${D^{\ast}}$ a (${D^{\ast}}- D$) mass difference)
consistent with the nominal mass \cite{PDG}.
The ${K^{\ast}}$, $\rho$ and $a_1^+$ masses are each required to be within
one full decay width of the nominal mass.

{}From the measured components, we obtain
a candidate momentum ${\bf{p}}_B=\sum{\bf p}_i$ and energy
$E_B=\sum{E_i}$.
We calculate $\delta (\Delta E) = {(E_{\mbox{beam}} - E_{B})/\sigma(\Delta
E)}$ $=$ $\Delta E /\sigma(\Delta E)$ and
the beam-constrained mass
$M_{B} = [{E_{\mbox{beam}}^2-(\sum{\bf p}_i)^2}]^{1/2}$,
where $E_{\mbox{beam}}$ is the beam energy
and $\sigma (\Delta E)$ is calculated
from the error matrices of daughter particles.
To be considered further a candidate must satisfy
${|\delta (\Delta E)|}<7.0$, $M_{B} > 5.2$~GeV.
We consider charged and neutral $B$'s separately, but all modes
are otherwise combined.
At most one charged and one neutral $B$ candidate per event are allowed;
for each, a ${\chi^{2}}$ is constructed based on the agreement
of its component ${\pi^{0}}$, ${K^0_{\rm S}}$, $D$, ${D^{\ast}}$ and
charged hadron candidates to their hypotheses, and the candidate with
the highest likelihood is selected.

Due to the spin orientation of the ${\Upsilon (4S)}$ produced in
${e^{+}}{e^{-}}$ annihilation,
the angle $\theta_B$ of the $B$ momentum with the beam axis
is distributed as $\sin^2\theta_B$.
We require $|\cos\theta_{B}|$ $<$ 0.95.
The ``thrust angle'' $\theta_{\mbox{thr}}$, between
the thrust axes of the $B$ candidate and of the remainder
of the event, is also examined.
The distribution in $\cos\theta_{\mbox{thr}}$ is uniform for $B$ decays and
peaks near  $|\cos\theta_{\mbox{thr}}|=1$ for continuum.
We require $|\cos\theta_{\mbox{thr}}|<0.9$~(0.8,0.7) for $B{\rightarrow} X\pi$
($B{\rightarrow}X\rho$, $B{\rightarrow}Xa_1$) where X is $D$ or ${D^{\ast}}$.

We define two regions in ${|\delta (\Delta E)|}$, ``signal''
(${|\delta (\Delta E)|}<2.5$) and ``sideband''
($4.0<{|\delta (\Delta E)|}<6.5$).
Looking both at data on the continuum and simulations of $B\bar B$ background,
the signal and sideband regions are found to give similar distributions
in $M_B$.
We require that the difference between $M_B$ and the nominal
$B$ mass be less than 6~MeV and count candidates, subtracting directly
the sideband from the signal sample.
We find $834\pm 42$ ${B^-}$ and $515\pm 31$ $\bar{B}^0$ mesons.
The same technique is used in events containing an identified lepton.
Figure~\ref{fig:allfig} (a) and (b) show the sideband superimposed on
the signal for both samples.
Details of this analysis are given in ref. \cite{mal}.

%
% ================KK, MSS analysis========================================
%
The method (B) of partial $\bar{B}^0{\rightarrow}{D^{\ast+}}
{\ell^{-}}{\bar{\nu}}$ reconstruction,
where the decay ${D^{\ast+}}{\rightarrow}{D^{0}}{\pi^{+}}$ is identified using
only the
${\pi^{+}}$,
has been used by CLEO II to measure $\bar{B}^0$ mixing \cite{mix}.
This approach yields a large sample and exploits
the extremely low energy of the ${D^{\ast+}}$ decay;
the momentum of the ${\pi^{+}}$ is scaled to obtain an approximate
four-momentum ($\widetilde{E}_{{D^{\ast}}},\widetilde{\bf{p}}_{{D^{\ast}}}$)
for
the ${D^{\ast+}}$.
The squared missing mass,
\begin{eqnarray*}
\widetilde{M}_\nu^2\equiv (E_{\mbox{beam}}-\widetilde{E}_{{D^{\ast}}} -
E_{\ell})^2-(\widetilde{\bf{p}}_{{D^{\ast}}}+
{\bf{p}}_{\ell})^2
\end{eqnarray*}
approximates the squared mass of the neutrino.

The pion and lepton momenta $p_{\pi}$ and $p_{\ell}$ are required to satisfy
$ p_{\pi} < 0.19\ {\rm GeV}$ and $1.8~<~p_{\ell}~<~2.4~{\rm GeV}$.
Pion candidates are
required to have a consistent specific ionization.
Background from charged $B$'s arising from
$B^-\rightarrow{D^{\ast\ast0}}{\ell}\nu$
(${D^{\ast\ast0}}\to{D^{\ast+}}{\pi^{-}}$)
is suppressed by demanding a high lepton momentum.

We examine the $\widetilde{M}_\nu^2$ distribution and select candidates in the
signal region ($\widetilde{M}_\nu^2>-2$~GeV$^2$).
The continuum contribution is estimated using the data collected at
energies off resonance, corrected for luminosity and energy differences.
Background from ${B\bar{B}}$ events is estimated via Monte Carlo.
Its shape in $\widetilde{M}_\nu^2$ is largely defined by the phase space, and
the simulation
is in agreement with data in the nonsignal regions and for
combinations where the lepton and pion carry the same charge.
Its normalization is obtained by fitting
to data in the sideband region, $-20<\widetilde{M}_\nu^2<-4$~GeV$^2$.
We find $7119\pm 139$ tags \cite{curlers}.
The same procedure is applied to tagged events with a lepton.
We require that
the cosine of the angle between the two leptons in these events
be less than 0.99, to eliminate multiply reconstructed tracks.
The $\widetilde{M}_\nu^2$ distributions obtained after continuum
subtraction are shown in Fig.~\ref{fig:allfig}, (c) and (d).

%=============================D* pi tags==============================
%
The decay chain $\bar{B}^0{\rightarrow}{D^{\ast+}}{\pi^{-}}$,
${D^{\ast+}}{\rightarrow}{D^{0}}{\pi^{+}}$ produces a hard
${\pi^{-}}$ and a soft ${\pi^{+}}$, which are used in method (C) to
identify the chain without detecting the ${D^{0}}$\cite{bcleo}.
Briefly, the energy ${E_D}$ of the ${D^{0}}$ is obtained by energy
conservation.
Since the ${D^{0}}$ and ${\pi^{+}}$ must form a ${D^{\ast+}}$,
defining their energies fixes the angle $\theta$ between them
as well as the decay angle ${\theta^{*}}$ of the ${\pi^{+}}$ in the
${D^{\ast+}}$ rest frame relative to the ${D^{\ast+}}$ direction in the lab.
If one then calculates the maximum possible $D^{\ast+}\pi^-$ invariant
mass $M_B$ under these constraints, it must for a
signal candidate lie in the narrow region
between the actual $B$ mass and the beam energy.
Details are given in ref. \cite{twobody}.

Taking events with $R_2<0.45$,
we select pairs of oppositely charged tracks
compatible with the tagging mode, i.e.,
for which values of ${E_{D}}$ and ${\cos\theta}$ are physical.
The fast track must have specific ionization consistent with being
a pion and not be identified as an electron or muon.
Because the ${D^{\ast+}}$ is fully polarized, the ${\pi^{+}}$ decay angle
is distributed as $\cos^{2}{\theta^{*}}$, and we require
$|{\cos{\theta^{*}}}|>0.5$.

Each event is partitioned approximately into
the tag $B$ and the other $B$ by
including with the candidate the two particles (charged
tracks or isolated neutral clusters) with the largest momentum component
opposite the fast pion.
The angle ${\theta_{\mbox{thr}}}$ is then calculated as in tag (A).
We require $|{\cos~{\theta_{\mbox{thr}}}}|<0.7$.

The distribution in $M_B$ is fitted to a Gaussian plus background.
The signal mean and width are obtained via Monte Carlo simulation.
We define a signal region ${M_{B}}> 5.276$~GeV and
a mass sideband ${M_{B}}< 5.26$~GeV.
Uncorrelated track pairs from ${B\bar{B}}$ events give a flat distribution.
Correlated pairs from
$B {\rightarrow}{D^{\ast\ast}}{\pi^{-}}$
(${D^{\ast\ast}}{\rightarrow}{D^{\ast+}}\pi$,
${{D^{\ast+}}{\rightarrow}D^{0} {\pi^{+}}}$)
produce a broad peak in the signal region.
The fitted background shape is the sum of this shape plus a first
order polynomial.
We find $822\pm 53$ tags.
The number of tagged events with an additional lepton is determined
by counting candidates in the signal region, subtracting
backgrounds from continuum, random ${B\bar{B}}$ combinations and
$B{\rightarrow}{D^{\ast\ast}}{\pi^{-}}$.
To exclude secondary leptons
from the undetected ${D^{0}}$, which are nearly opposite
to the fast pion, we require the cosine of the angle
between the fast pion and lepton to be greater than $-0.85$.

%
%=============================Lepton Selection========================
%
For all three tag types, additional leptons are selected
in the range $1.4-2.4$~GeV.
Background occurs when the reconstructed
decay is correct but the lepton is fake (a hadron passing lepton
identification) or, for neutral $B$'s, a secondary from
charmed meson decay.
Fakes are assessed separately for each type of tag.
In each case, events which contain candidates in
the signal regions are selected.
Excluding those which comprise each candidate,
all tracks within the acceptances of lepton identification
which fail the identification criteria are considered.
These tracks, each weighted by
the appropriate fake rate per track, are summed.
The contribution from background candidates is estimated
using events selected from the various sidebands.
The fake rate as a function of momentum
is determined using data taken at the ${\Upsilon (1S)}$, which produces very
few leptons.
Of all leptons from $B$ decay with momentum above 1.4~GeV,
the fraction from secondaries is found to be $0.028 \pm 0.008$.
The fraction among detected
leptons is $0.027 \pm 0.008 (0.022\pm 0.007)$ for electrons (muons).
For tags (B), these
factors are twice as large because the undetected ${D^{0}}$ may also
contribute.
These corrections are applied to the neutral $B$ tags.

The efficiency for geometric acceptance, track reconstruction
and identification is found to be 65.1\% (50.7\%) for electrons (muons).
Requirements on dilepton opening angle and event characteristics
result in effective adjustmentments of
$-1.7\%$, $-1.1\%$ and $-12\%$ for (A), (B) and (C), respectively.
Dependences of tagging efficiencies on the decay charge multiplicity of
the other $B$ in an event result in effective increases
of 6\% for (A) and 3\% for (B) and (C).
To extrapolate the spectrum to lower momenta,
we use the model of Isgur et al. ($ISGW$),\cite{lepmodel} which
predicts that three exclusive modes dominate, $B{\rightarrow}{D}{\ell}\nu$,
$B{\rightarrow}{D^{\ast}}{\ell}\nu$ and
$B{\rightarrow}{D^{\ast\ast}}{\ell}\nu$.
Based on our fit to the inclusive lepton
spectrum,\cite{rwang} we take their proportions to be 24.5/54.5/21.
For electrons (muons) the fraction of the spectrum above
1.4~GeV is found to be 48.1\% (51.4\%).
Assuming $e-\mu$ universality, we average the electron and muon totals.
Shown in Table I are the raw numbers of tags with and without
leptons, corrections, and branching fractions.

%
%========================Systematic Errors=============================
%
The systematic uncertainties in efficiencies
for tracking, lepton identification and lepton spectrum extrapolation
are common to all of the analyses and are found to be 2.0\%, 2.5\%
and 8\%, respectively.
The effect of the lepton spectrum is simulated by varying the percentage
of $B{\rightarrow}{D^{\ast\ast}}{\ell}\nu$ from 0\% to 30\% in the ISGW model.
Uncertainties due to fitting methods and functions
are determined by varying the techniques and shapes used to
fit signals and backgrounds.
Selection of events and candidates add uncertainties
which are estimated via Monte Carlo.
The uncorrelated systematic uncertainties  are summarized in Table~II.

We average the $\bar{B}^0$ measurements, using the quadratic sum of
statistical and uncorrelated systematic errors to determine
relative weights.
The results are
\begin{eqnarray*}
b_- = (10.1 \pm 1.8\pm 1.4)\%,\\
b_0 = (10.9 \pm 0.7\pm 1.1)\%.
\end{eqnarray*}
Both are consistent with the CLEO II average inclusive
branching fraction, $(10.9\pm 0.1\pm 0.3)\%$ \cite{rwang}.
The ratio is
\begin{eqnarray*}
{b_-\over b_0}= 0.93 \pm 0.18 \pm 0.12.
\end{eqnarray*}
The error due to uncertainty in ${f_{+-}}/{f_{00}}$ is less than 1\%.
This result is in agreement with our other result
and with that of ARGUS \cite{RMEASURED}, which
assume that ${{f_{+-}}/{f_{00}}}=1$.
It is consistent with the world average lifetime ratio of
$1.10\pm 0.11$ \cite{LEP}
as well as with theoretical expectations, and is of comparable significance to
existing individual measurements.

We gratefully acknowledge the effort of the CESR staff in providing us with
excellent luminosity and running conditions.
This work was supported by the National Science Foundation,
the U.S. Dept. of Energy,
the Heisenberg Foundation,
the  SSC Fellowship program of TNRLC,
and the A.P. Sloan Foundation.

%
%=============================References===============================
%

%**************************tables

\begin{table}
%\begin{center}
\caption{Numbers of tags without ({all}) and
with ($e,\mu$) additional leptons.
Tag types are described in the text.
Subtraction of estimated fakes (a) and secondaries (b) yields
the number of detected primary leptons (c), which is corrected
for efficiency to obtain (d).
The electron and muon values are averaged to obtain $N_{\ell}$,
and the branching fraction {\cal B}.}
\begin{tabular}{  l c c c c }
& ${B^-}$ (A) & $\bar{B}^0$ (A) & $\bar{B}^0$ (B) &
$\bar{B}^0$ (C) \\
\hline
{all}       &$834  \pm 42  $&$515 \pm 31  $&$7119\pm 139$ &$822  \pm 53  $\\
\hline
$e$       &$ 32.0\pm  6.8$& $23.0\pm  5.3$ &$271.9\pm 22.1$ &$ 25.8\pm  6.2$\\
(a) &$ 0.17\pm 0.05$&$0.13\pm 0.04$ &$1.5\pm 0.4$ &$ 0.1\pm 0.02$\\
(b) &---&$ 0.7\pm 0.2$ &$14.6\pm 4.3$ &$0.7\pm 0.2$\\
(c) &31.8$\pm $6.8& $22.3 \pm$5.3 &$255.8\pm 22.5$ &25.0$\pm $6.2\\
(d) & $97.2\pm 20.8$& $68.0\pm 16.2$
  &$826.0\pm 72.7$ &$90.9\pm 22.5$\\
\hline
$\mu$      &$ 21.0\pm  5.4$&$ 21.0\pm  5.4$ &$197.7\pm 20.2$ &$ 19.6\pm  5.5$\\
(a) &$  1.1\pm  0.3$&$ 0.9\pm  0.3$ &$8.3\pm 2.5$ &$  0.5\pm  0.1$\\
(b) &---&0.5$\pm$0.2 &$8.3\pm 2.7$ &$0.4\pm 0.1$\\
(c) &19.9$\pm$5.4& 19.7$\pm$5.4
 &$181.1\pm 20.5$ &18.7$\pm$5.5\\
(d) &$73.0\pm 19.8$& $72.1\pm 19.9$
 &$702.5\pm 79.6$ &$81.7\pm 24.0$\\
\hline
$N_{\ell}$ &$84.6\pm 14.3$& $69.6\pm 12.5$
 &$767.5\pm 53.5$ &$86.6\pm 16.3$\\
\hline
{\cal B}& $10.1 \pm 1.8 $& $13.5\pm 2.6 $
 &$10.5\pm 0.8$ &$10.2 \pm 1.9 $\\
\end{tabular}
%\end{center}
\end{table}

\begin{table}
\caption{Uncorrelated systematic errors,
given as percent of signal.}
\begin{tabular} {  l c c c c }
& ${B^-}$ (A) & $\bar{B}^0$ (A) & $\bar{B}^0$ (B) &
$\bar{B}^0$ (C) \\
\hline
Event selection       & 1.7 & 1.7 & 0.5  & 4.1  \\
Single tag/event      & 7.2 & 5.0 & --   & --   \\
\# tags               & 4.8 & 6.5 & 2.6 & 4.6 \\
\# tags with leptons  & 3.4 & 5.1 & 7.1 & 5.6 \\
Tagging efficiency    & 6.0 & 6.0 & 3.0 & 3.0 \\
Secondary leptons     & --  & 0.8 & 1.6 & 0.8 \\
Fake leptons          & 0.1 & 0.2 & 0.2 & 0.2  \\
\hline
Total uncorrelated    &11.2 &11.5 & 8.3 & 8.9 \\
\end{tabular}
\end{table}

%***********************figures FOR DRAFT

\begin{figure}
\vspace{12. cm}
\includegraphics{full.ps}
\vspace{9. cm}
\includegraphics{part.ps}
 \caption{Tag samples, without and with additional
lepton required.
(a), (b): $B^-$, method (A), $M_B$ distributions;
(c), (d): $\bar B^0$, method (B), $\widetilde{M}_\nu^2$
distributions.}
 \label{fig:allfig}
\end{figure}


\begin{references}
\bibitem{cconj}
Throughout this paper, mention
of a specific particle and decay mode implies also its charge conjugate.
\bibitem{curlers}
About 5\% of tags are found to be identified more than once because
the soft pion curls and is multiply reconstructed.
We account for this by increasing our statistical error by 2.5\%.
\end{references}
\end{document}